%% file: d2d_offloading1_WPMC_v2.tex
\documentclass[a4paper, conference, 10pt]{IEEEtran}
\usepackage{verbatim}
\usepackage[dvips]{graphics} 
\usepackage{graphicx}
\usepackage{times}
\usepackage{epsfig}
\usepackage{latexsym}
\usepackage{amsmath}
\usepackage{cite}
\usepackage{textcomp}
\usepackage{enumerate}

\usepackage{authblk}\usepackage{algorithmic,algorithm}
\begin{document}

\title{Efficient Device-to-Device (D2D) Offloading Mechanism in LTE Networks}
 
\author{Bighnaraj Panigrahi, Rashmi Ramamohan, Hemant Kumar Rath, Anantha Simha}
\affil{CTO Networks Lab, Tata Consultancy Services, Bangalore, India, \\Email:\{bighnaraj.panigrahi, rashmi.ramamohan, hemant.rath, anantha.simha\}@tcs.com}
\maketitle

\begin{abstract}
With the increase in mobile traffic and the bandwidth demand, Device-to-Device (D2D) communication underlaying Long Term Evolution (LTE) networks has gained tremendous interest by the researchers, cellular operators and equipment manufacturers. However, the application of D2D communication has been limited to emergency services and it needs to be explored in commercial applications. In this paper, we have introduced a novel commercial D2D offloading scheme for users who may be at cell edges, inside isolated environments like basement or large buildings, etc. Our proposed scheme discovers the available idle D2D neighbors for such poor channel users and offloads its data to the D2D neighbor, which then relays the data to the eNB. We have developed a D2D offloading simulation model in MATLAB, have conducted extensive simulations and have observed that the proposed scheme can provide better efficiency to the network as well as satisfy the poor channel users significantly.
\end{abstract}


\section{Introduction}\label{sec:intro}
There is a huge demand for high data rate services like video sharing, online gaming, proximity-aware-social-networking, etc., in today's cellular network (2G/3G/4G). However, the vulnerability of the cellular networks significantly affects the performance of these services; not only hampers the data communication, but also adds  unfairness to network usage of the users. Because of the vulnerability, cell edge users, users inside a lift or basement, etc., do not get a fair Quality of Services (QoS) experience as compared to other users in the network. Studies have been conducted by various researchers in this direction so as to ensure high data rates to poor channel users and to maximally utilize the wireless resources.

Today, Device-to-Device (D2D) communication \cite{Doppler2009, Panigrahi2015}, in which close proximity User Equipments (UEs) can directly communicate among themselves using the cellular resources and bypassing the Evolved NodeB (eNB) is getting focus under Long Term Evolution (LTE) networks. In D2D communication, UEs directly can communicate under the control and assistance of the eNB. This helps in direct transfer of data between two close-by UEs at a higher rate with lower transmission power and delay. D2D can also be applied for many other applications; one such use-case is using D2D communications for opportunistic offload of poor channel users' data. Under opportunistic offload, whenever a poor channel user gets an idle close-proximity neighbor UE with a good channel profile to the eNB, it can offload its data to the neighbor UE using D2D communication and then the neighbor UE can relay the data to the eNB using regular cellular communication. Though this is a two hop communication, it provides higher 
level of assurance on the data delivery and is spectrally efficient. This kind of offloading can also be used for multi-hop relays in cellular networks, multi-casting \cite{Seppala2011}, peer-to-peer communication, video dissemination, Machine-to-Machine (M2M) communication, and so on. We believe that cellular operators and the mobile users will be greatly benefited by D2D offloading and will show interest to incorporate the changes related to it.

D2D offloading requires network assistance. Therefore, appropriate modifications are required in terms of signaling, scheduling, and synchronization by the eNB. This also requires the knowledge of the location information of the UEs and channel conditions between UE-UE and UE-eNB transmissions. The location information of the UEs can be availed using the control messages and by the possible use of Global Positioning System (GPS) or any other efficient localization techniques. The eNB needs to identify if any neighbor of the poor channel UE is eligible and willing to be the D2D offloader. Once the eNB identifies the D2D offloader, it allocates resources for D2D offloading (UE-UE) and for the cellular relay (UE-eNB) serially so that UE-eNB relaying happens only after the UE-UE offloading.
 
In this paper, we have introduced a novel D2D offloading scheme and have extended the 3rd Generation Partnership Project (3GPP) proposed method for discovery in Proximity Services (ProSe) \cite{3GPP2012, Lin2014} using an optimization framework. We have also proposed a 2-phase scheduling scheme to schedule D2D as well as cellular communication. The rest of this paper is organized as follows. In Section \ref{s:relWork}, we provide a brief literature survey on D2D offloading. Section \ref{s:system_model} describes the system model and formulation of the problem. In Section \ref{ss:offload}, we describe the offloading process, in Section \ref{s:sec_expt}, we evaluate the performance of our scheme and conclude in Section \ref{s:sec_conclusion}.

\section{Related Works}{\label{s:relWork}}
The use of cellular spectrum resources in uplink/downlink for the D2D communication is mentioned in \cite{Doppler2009, Panigrahi2015}. It is also proposed as a form of relaying \cite{Wu2001}. Radio aspects of D2D discovery and communication was submitted by Qualcomm to 3GPP in 2011 \cite{3GPP2012}. The main focus of 3GPP is to design a LTE-enabled proximity service for the emergency conditions wherein two devices can communicate directly to each other without the intervention of the operator when the network is non-functioning. Offering D2D services as a commercial service has not gained much attention due to the standardization, billing, regulatory and security related issues. In commercial case, the D2D users have to be guaranteed with better QoS than the traditional technologies such as Bluetooth, Infrared, etc., which are being offered at no cost but with lower speed and compromised security. Data offloading has been considered as one of the key solutions to handle the traffic on the networks by 3GPP. 

In \cite{Sankaran2012}, the basic concepts of offloading methods in Internet Protocol (IP) networks are discussed. Offloading the cellular traffic to the WiFi networks are studied in \cite{Deep2014, Kyunghan2012}. In a similar way, studies have also been carried out in offloading the traffic to small cells or femto cells \cite{3GPP23.829}. The authors in \cite{Alexander2013} have explored offloading the cellular traffic onto the WiFi Direct links and have analyzed their performance with respect to the LTE networks. Assisted offloading has been presented in \cite{Sergey2015} where the authors have modeled the dynamic network and demonstrated the offloading process by assuming cellular users in licensed spectrum and the D2D users in the unlicensed spectrum. However, these data offloading schemes are different from D2D offloading as they follow a different network architecture; availability and reliability can not be guaranteed. 

In this paper, we have proposed a novel optimal D2D offloading scheduling scheme for users with unfavorable channel conditions to eNB in LTE-based cellular networks.

\section{System Model and Problem Formulation} \label{s:system_model}
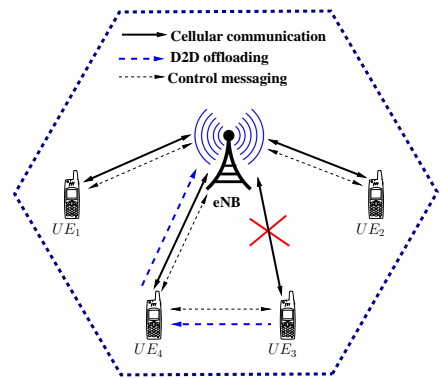
\begin{figure}[h!]
\centering 
\scalebox{0.4}{\input{./D2Dframeworknew.pstex_t}}
\caption{Single Cell D2D Offloading}  
\label{topology}\vspace{-3mm}
\end{figure}

We consider a single cell with one eNB and multiple UEs as shown in Fig. \ref{topology}. The UEs are assumed to be static (full-fledged mobility of the UEs is out of scope of this paper) so that there will be minimum fluctuations in the channel conditions. The network is assumed to have technologies in place to accommodate the D2D communication with all the necessary synchronization and control mechanisms. We further assume that the following two types of communications are possible: (i) Cellular communication where the UE directly transmits to the eNB and (ii) D2D offloading where UE relays data to the eNB via another close-by UE. Based on the wireless channel condition between the UEs and the eNB, we divide the UEs into two sets: $Type-1$ users - who have a good channel to the eNB and $Type-2$ users - who have a poor channel to the eNB. Note that, while the $Type-1$ users communicate using cellular mode, $Type-2$ users use D2D offloading communication. In a typical scenario as shown in Fig. \ref{topology}, 
$UE_3$ has a poor channel to the eNB, can offload its data to the close-by user $UE_4$ to be ultimately relayed to the eNB. Once eNB receives the request of the data offload for $UE_3$, it initiates the D2D offload process along with necessary scheduling and controlling methods. We further assume that the users transmit at fixed rates between the UE to eNB and UE to UE using fixed modulation scheme. However, the same model can be extended to the case of variable data rate and modulation schemes. The transmission power of the UEs can be controlled by the eNB, such that the received Signal to Noise Ratio ($SNR$) is above a specified threshold ($SNR_{th}$). 

\subsection{Frame Structure}
\begin{figure}[h!]
\centering 
\includegraphics[width=0.4\textwidth]{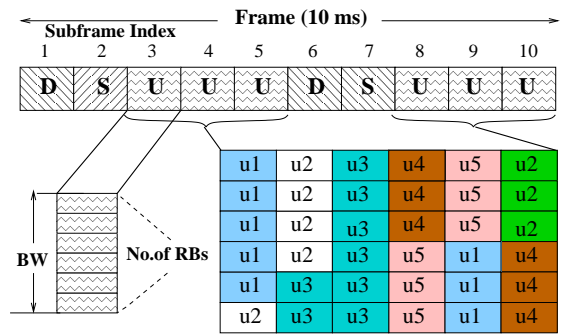}
\caption{LTE Uplink TDD Frame Structure}  
\label{frame_FDD}\vspace{-3mm}
\end{figure}

We consider standard LTE Time Division Duplex (TDD) frame structure - periodicity: 5 ms, frame length: 10 ms and each frame consists of 10 sub-frames (1 ms each). The sub-frames can either be uplink (U), downlink (D) or special (S). As shown in the Fig. \ref{frame_FDD}, six sub-frames are used for uplink scheduling. We assume that the system bandwidth is 1.4 MHz and consists of six Resource Blocks (RBs); bandwidth of each RB is 180 kHz. Each RB consists of 12 sub carriers. For D2D communications uplink frames are only preferred.


We also assume that the UEs have different amount of uplink data to transmit to the eNB. With fixed modulation scheme and fixed data rate in place, the eNB can compute the number of RBs required by each UE before hand and can schedule the number of RBs in a frame by frame basis. Note that, the performance of the network depends upon the selection of the mode of communication, i.e., using direct cellular or D2D offloading and the selection of the offloader in the possible case when multiple neighbors can act as an offloader for a particular UE. With multiple D2D offloaders available, the eNB has to choose the offloader which ensures optimal throughput as discussed below.

\subsection{Optimal Mode and Offloader Selection} \label{ss:mode}
Let $m$ be the number of D2D offloaders available for $UE_i$. This can be feasible in a real scenario by scanning for D2D offloaders within a small D2D radius of $r_d$. Now, $UE_i$ has one direct (cellular) and $m$ offloading paths to choose from. Let $x_{ij}, \forall j \in (1,...,m, e)$ be a Boolean variable which indicates whether the communication between node $i$ and $j$ is active or not (symbol $e$ indicates eNB). For example, $x_{ie} = 0$ indicates that there is no direct path from node $i$ to eNB and $x_{ij} = 0$ indicates that node $j$ can not be an offloader for $UE_i$ because of unavailability, mobility, etc., of the offloader. Let $p_{ij}$ be the probability of an unsuccessful transmission of an RB between node $i$ and $j$. Expected number of Hybrid Automatic Repeat reQuest (HARQ) retransmissions required per RB between node $i$ and $j$ can be given as $r_{ij} = \dfrac{1}{p_{ij}}$. Let $k_{ij}$ be the number of RBs required to transmit the desired data from node $i$ to $j$ (computed by eNB as 
assumed earlier) and $c_{ij}$ is the cost associated for each RB transmission. Then, the total cost associated with the transmission between node $i$ and $j$ is:


\begin{equation}\label{eq:c}
	\begin{aligned}
	&&& \ \ C_{ij} = \ \ k_{ij} \times \ r_{ij}  \times \ c_{ij}. 
	\end{aligned}
	\vspace{-2mm}
\end{equation}

In addition to the above, $UE_j$ also needs to relay the data received from $UE_i$ to eNB. The cost associated with this relay can also be calculated as in Eqn. (\ref{eq:c}). Let the cost associated with this relay be $C_{je}$ and $C_{ie}$ be the cost associated with the direct transmission between $UE_i$ and eNB. Using the above costs for transmission, eNB solves the following optimization problem such that the communication mode (direct or via offloader) and the optimal D2D offloader can be selected. 

\begin{equation}\label{eq:opt}
	 \begin{aligned}
	&&& \min \ \LARGE( C_{ie} \times x_{ie} + \sum_{j=1}^{m} (C_{ij}+C_{je}) \times x_{ij} \LARGE),  & &\text{s.t.} \\
	&&& (i) \  SNR_{ij} \ge SNR_{th} :\hspace*{2 mm} \forall j \in (1,..,m),  \\
	&&& (ii) \  SNR_{ie}, SNR_{je}  \ge SNR_{th} :\hspace*{2 mm} \forall j \in (1,..,m),  \\
	&&& (iii) \  x_{ie}, x_{ij} = 0/1  :\hspace*{2 mm} \forall j,  \\
	&&& (iv) \ x_{ie} + \sum_{j=1}^{m} \ x_{ij} \le 1  :\hspace*{2 mm} \forall  j. 
	\end{aligned}
	\vspace{-2mm}
\end{equation}

Constraints (i), (ii) of Eqn. (\ref{eq:opt}) ensure the SNR profiles for communications. Constraint (iii) defines the Boolean variables for the possibility of a path, and constraint (iv) ensures that out of ($m+1$) available paths, eNB selects only one. We now discuss the D2D offloading process as a solution to the above mathematical formulation.

\section{Offloading Process and Channel Model} \label{ss:offload}
We now explain the five-phase offloading process and power control mechanism that needs to be used for the successful data transmission. 

\subsection{Offloading Phases}

\subsubsection{Offloading Initiation Phase} \label{sss:discovery}
Offloading initiation can either be made by the eNB or by the UE. In the eNB initiated type, eNB offers D2D offloading option depending on the UE's location and availability of D2D offloader. In UE initiated type, if the UE with a poor channel can send the control messages (usually transmitted at a higher power) to the eNB, then the eNB can make the necessary arrangements for D2D offloading. In case the UE is in a completely isolated environment, it can relay the offloading initiation request through a known close-by UE who has a good channel to the eNB. 

\subsubsection{Offloader Discovery Phase} \label{sss:discovery}
In this phase, optimal offloader is discovered as a solution to Eqn. (\ref{eq:opt}) which can be either network assisted or ad-hoc type. In network assisted type, the channel condition between the UE ($UE_3$ in Fig. \ref{topology}) and the eNB is such that only offloading request can be communicated through control channels. The eNB which is capable of extracting information of the UE, its neighbors' locations ($UE_3$ and its neighbors in example), their load and channel conditions, etc., can select a suitable offloader ($UE_4$ for $UE_3$). In the ad-hoc discovery type, the $Type-2$ UE itself finds a suitable offloader among its nearest neighbors with a request/reply method and then conveys this to the eNB. In this case, the part of the optimization problem Eqn. (\ref{eq:opt}) is solved by the UE itself. Although, network assisted discovery is more power consuming, it is reliable and backward compatible. 


\subsubsection{Handshaking Phase} \label{sss:handshake}
In this phase, control messages are exchanged between the $Type-2$ UE and eNB and between eNB and D2D offloader. After the finalization of the D2D offloader, eNB collects the required information like availability, channel condition, etc., and exchanges the synchronization information with the offloader UE. The handshaking mechanism is diagrammatically shown in Fig. \ref{flowdiagram}. 

\subsubsection{Scheduling Phase} \label{sss:schedule}
For data transmission, UE sends Scheduling Request (SR) signaling in the Physical Uplink Control Channel (PUCCH) to the eNB. The eNB then assigns RBs to the requester and the offloader. We propose a two-phase scheduling scheme; in the first phase, D2D transmission is scheduled and in the second phase the transmission of D2D offloader to the eNB is scheduled. These two phases can take place in same or in different LTE time frames without any time overlapping of a Phase-1 RB with any Phase-2 RB. The scheduling can be overlapping or non-overlapping type \cite{Panigrahi2015}; we have assumed non-overlapping type of scheduling in this paper, i.e., both cellular and D2D communication will use non-overlapped RBs for their communication. 

\subsubsection{Communication Phase} \label{sss:comm}
\begin{figure}[!htb]
\centering
\includegraphics[width=0.45\textwidth]{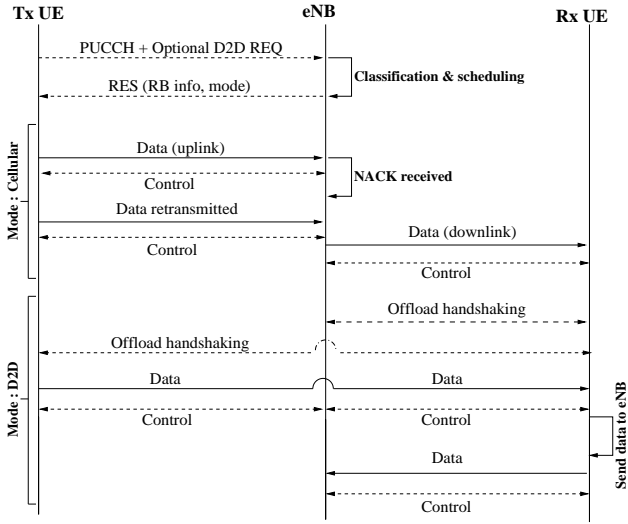}
\vspace{-3mm}
\caption{Control Flow Diagram \label{flowdiagram}}
\end{figure}

This is the actual data communication phase between D2D transmitter and offloader followed by between offloader and eNB. The control flow diagram is illustrated in Fig. \ref{flowdiagram} and the corresponding algorithm for offloading process is described in Algorithm \ref{offloadAlg1}. Note that, it is not necessary for all five phases to be executed serially.

\begin{algorithm}[h] \smallskip \small
\caption{: D2D Offloading Algorithm}
\label{offloadAlg1}
\begin{algorithmic}[1]
\STATE eNB receives SR signal from UEs: \{location, No. of RBs required, channel condition\}\\
\COMMENT {Offloading Initiation Phase}
\STATE eNB computes $SNR_{ie}$ \\
\COMMENT {Offloader Discovery Phase}
\IF {$SNR_{ie} > SNR_{th}$}
  \STATE $UE_i$ is a $Type-1$ user; Go to Line 8 
\ELSE 
  \STATE Find the D2D offloader as a solution to Eqn. (\ref{eq:opt}). 
\ENDIF \\
\COMMENT {Handshaking and Scheduling phase}
\STATE Phase-I: Schedule the $Type-1$ cellular and D2D offload (UE-UE) transmissions 
\STATE Indicate the RBs, transmission power and modulation scheme to be used by the UE
\STATE Phase-II: Schedule the D2D relay (UE-eNB) transmissions 
\STATE Indicate the RBs, transmission power and modulation scheme to be used by the UE \\
\COMMENT {Communication Phase}
\STATE Starting of actual communication
\end{algorithmic}
\end{algorithm}
\vspace{-4mm}

\subsection{Channel Model} \label{ss:ch}
The channels between the UE-eNB and UE-UE are assumed to be remain constant for at least one LTE frame duration. We assume that the channel variations follow a combination of log-normal shadowing and multi-path Rayleigh fading in addition to the variations due to distance and other factors. Let $Pt_{max}$ be the maximum transmit power in dBm, $d_{ij}$ be the distance between the $UE_i$ and node $j$ ($j$: eNB/UE). Received power $Pr_{ij}$ can be expressed in logarithmic form as:  


\vspace{-2mm}
\begin{equation}\label{eq:pow_recd}
	\begin{aligned}
	&&& \ \ Pr_{ij}=Pt_{max} - P^{loss}_{ij}, \\
	&&& \ \ P^{loss}_{ij}= {\gamma log_{10} (d_{ij}+f)} + {L_{sh}} + {L_{rl}} + {L_{a}}, \\	
	\end{aligned}
	\vspace{-2mm}
\end{equation}

\noindent where, $f$ is the frequency of operation, and $\gamma$ is the path loss exponent, $L_{sh}$ and $L_{rl}$ be the losses due to the Log normal shadowing and Rayleigh fading respectively and ${L_{a}}$ is the location specific loss where the user is placed. Location specific loss should be factored in the path loss computation as the $Type-2$ user can be placed inside a lift or in the basement or some other place similar to this. We assume that the eNB can compute the path loss parameters and the total loss value $P^{loss}_{ij}$. If $N_o$ is the Additive White Gaussian Noise power, then instantaneous $SNR_{ij} = \dfrac{Pr_{ij}}{N_{o}}$. For successful data transmission, $SNR_{ij} \ge SNR_{th}$. 

\subsubsection{Power Control for D2D Communication} \label{sss:power}
Optimal transmission power required for D2D communication should be computed by the eNB as follows and should be informed to the UE in the handshaking phase. Since, it is difficult to have continuous values for $Pt_{c_{ij}}$, we have defined discrete levels for $Pt_{c_{ij}}$: [-5, 5, 15] mw and select the appropriate level based on the distance and $SNR_{th}$ condition. 

\begin{equation}\label{eq:red_tx_pow}
	 \begin{aligned}
 	 &&& \ \ Pt_{c_{ij}} \ge \ \ SNR_{th} \times N_{o}  + P^{loss}_{ij}.
	 \end{aligned}
\end{equation}

\section{Experimental Evaluation}\label{s:sec_expt}
We have considered a single eNB with 100 UEs placed at random positions within a radius of 150 m. On an average maximum 50 users are assumed to be ready for transmission and termed as active transmitter UEs. The remaining users are considered to be idle and can be used as potential D2D offloaders. In order to realize a practical scenario, we have assumed on an average $40\%$ of the active transmitter UEs as $Type-2$ users. The received power of each UE is calculated using Eqn. (\ref{eq:pow_recd}). We have considered two different scenarios: (a) cellular only communication and (b) cellular with D2D offloading. We have also implemented a simple Round Robin uplink scheduling scheme. The offloaders are assumed to have sufficient battery power. The other simulation parameters used are mentioned in Table \ref{t:parameters}. 

\begin{table}[!tbh]
\caption{\small \label{t:parameters}Simulation Parameters}
\centering
\begin{tabular}{|l|l|}\hline
{\bf Parameter} & {\bf Values} \\ \hline
Tx Power and Range &  $23$ dBm, $150$ m \\ \hline
Path loss component ($\gamma$) & $3$ \\ \hline
Shadowing standard deviation($\sigma$) & $6$ dB \\ \hline
Frequency of operation (f) & $2300$ MHz \\ \hline
Modulation & 16 QAM \\ \hline
Maximum D2D distance & $20$ m \\ \hline
$SNR$ threshold ($SNR_{th}$) & $10$ dB \\ \hline
\end{tabular}
\end{table}
 \vspace{-5mm}
\subsection{Simulation Results}
\begin{figure}[!htb]
\begin{center}
\includegraphics[width=0.44\textwidth]{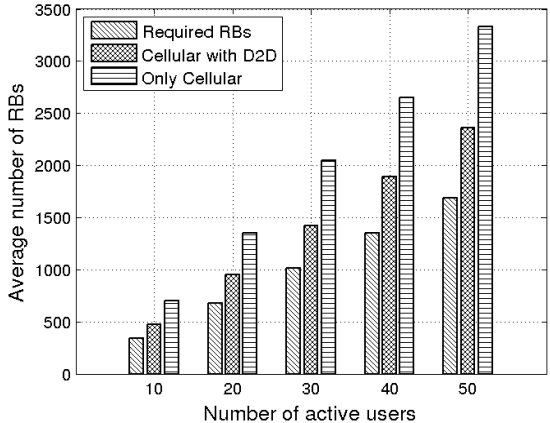}
\vspace{-3mm}\end{center}
\caption{\small\label{f:avg_bw_util} Average RB utilization vs. Number of active users}
\end{figure}

\begin{figure}[!htb]
\begin{center}
\includegraphics[width=0.44\textwidth]{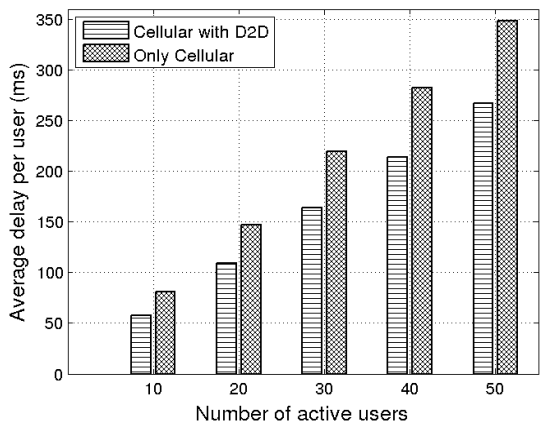}
\vspace{-3mm}\end{center}
\caption{\small\label{f:delay_per_user} Average delay per user vs. Number of active users}
\end{figure}

Sufficient simulation runs with different UE populations have been conducted with 100 different seed values to have results with a confidence interval of $\pm 2\%$ of the mean value with a confidence level of $95\%$. Fig. \ref{f:avg_bw_util} explains the average RBs utilization with varying number of active transmitters. From this figure, we observe that the average number of RBs consumed in case of D2D mode are much lower as compared to the conventional cellular mode; $40\%$ of RBs can be saved on an average results in significant improvement in spectral efficiency. This is due to the fact that multiple retransmissions have occurred in regular cellular communications as the channel between the UE and the eNB is bad. For the very same reason the average per user delay is also less for the D2D offload case (c.f. Fig.\ref{f:delay_per_user}); improves QoS of the users.

\begin{figure}[!htb]
\begin{center}
\includegraphics[width=0.44\textwidth]{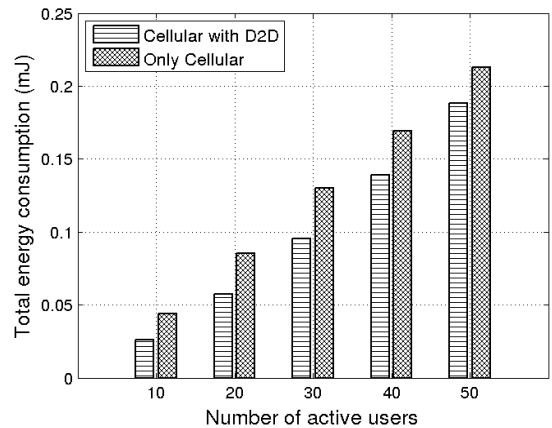}
\vspace{-3mm}\end{center}
\caption{\small\label{f:total_energy_con} Total energy consumption vs. Number of active users}
\end{figure}

From Fig. \ref{f:total_energy_con}, we observe that the total energy consumption is higher in case of regular cellular mode as compared to the D2D mode. This is due to the lower transmit power usage of the UEs and higher success rate of the D2D offloading as compared to the regular ones. Energy consumption is computed by considering the maximum transmission power of UE and the transmission time interval (TTI) of an RB. Energy performance also includes the overhead energy consumption in the D2D offloader discovery, handshaking, and control signal transmissions. System performance shown here in terms of RB utilization, delay and energy consumption only considers the Medium Access Control (MAC) level performance; further improvement is possible when higher layer parameters are taken into consideration.   

\vspace{-4mm}
\vspace{-4mm}
\vspace{-2mm}
\section{Conclusion}\label{s:sec_conclusion}
In this paper, we have proposed a D2D offloading mechanism to benefit the users who are experiencing a poor channel to the eNB. We have discussed an optimization framework which selects an appropriate communication mode for uplink and also the optimal offloader in case of D2D offloading. We have also proposed a 2-phase scheduling algorithm for D2D offloading. The proposed scheme is evaluated through extensive Matlab simulations. Our simulation results exhibit that with D2D offloading technique embedded in cellular network, network performance can be improved significantly. 

Implementing D2D offloading in LTE cellular networks requires appropriate signaling techniques. Moreover, the eNB should be able to compute the expected received signal strength at any UE location. Even at device level the UEs should be capable of changing their transmission powers as well as collecting their location informations. These modifications should be incorporated in the future LTE releases and should be standardized under 3GPP such that D2D offloading can be made possible seamlessly. The standardization effort should also encourage operators and vendors to adopt D2D implementation in future. 

\bibliographystyle{IEEEtran}
\bibliography{allRefOffloading}
\end{document}

%% file: D2Dframeworknew.pstex_t
\begin{picture}(0,0)%
\includegraphics{D2Dframeworknew.pstex}%
\end{picture}%
\setlength{\unitlength}{4144sp}%
\begingroup\makeatletter\ifx\SetFigFont\undefined%
\gdef\SetFigFont#1#2#3#4#5{%
  \reset@font\fontsize{#1}{#2pt}%
  \fontfamily{#3}\fontseries{#4}\fontshape{#5}%
  \selectfont}%
\fi\endgroup%
\begin{picture}(6340,5516)(1744,-5901)
\put(2341,-3751){\makebox(0,0)[lb]{\smash{{\SetFigFont{14}{16.8}{\rmdefault}{\bfdefault}{\updefault}$UE_1$}}}}
\put(5536,-5551){\makebox(0,0)[lb]{\smash{{\SetFigFont{14}{16.8}{\rmdefault}{\bfdefault}{\updefault}$UE_3$}}}}
\put(6841,-3751){\makebox(0,0)[lb]{\smash{{\SetFigFont{14}{16.8}{\rmdefault}{\bfdefault}{\updefault}$UE_2$}}}}
\put(4726,-3346){\makebox(0,0)[lb]{\smash{{\SetFigFont{14}{16.8}{\rmdefault}{\bfdefault}{\updefault}eNB}}}}
\put(3556,-5551){\makebox(0,0)[lb]{\smash{{\SetFigFont{14}{16.8}{\rmdefault}{\bfdefault}{\updefault}$UE_4$}}}}
\end{picture}%